\documentclass{amsproc}

\newtheorem{theorem}{Theorem}[section]

\newtheorem{proposition}[theorem]{Proposition}

\theoremstyle{definition}

\newtheorem{examples}[theorem]{Examples}

\theoremstyle{remark}

\numberwithin{equation}{section}

\begin{document}

\title[Approximations of vertex couplings in quantum graphs]
{Approximations of permutation-symmetric vertex couplings in quantum graphs}

\author{Pavel Exner}
\address{Department of Theoretical Physics, Nuclear Physics
Institute, Academy of Sciences, 25068 \v{R}e\v{z} near Prague,
Czechia} \email{exner@ujf.cas.cz}
%\curraddr{}
\thanks{The participation of the first author in the conference was
made possible by the NSF. A partial support of the research by
ASCR and its Grant Agency within the projects IRP AV0Z10480505 and
A100480501 is also gratefully acknowledged.}

\author{Ond\v{r}ej Turek}
\address{Department of Mathematics, Faculty of Nuclear Sciences and
Physical Engineering, Czech Technical University, Trojanova 13,
12000 Prague, Czechia} \email{oturek@centrum.cz}

%    General info
\date{September 30, 2005}



\begin{abstract}
We consider boundary conditions at the vertex of a star graph
which make Schr\"odinger operators on the graph self-adjoint, in
particular,  the two-parameter family of such conditions invariant
with respect to permutations of graph edges. It is proved that the
corresponding operators can be approximated in the norm-resolvent
sense by elements of another Schr\"odinger operator family on the
same graph in which the $\delta$ coupling is imposed at the vertex
and an additional point interaction is placed at each edge
provided the coupling parameters are properly chosen.
\end{abstract}

\maketitle

\section{Introduction}

There is no necessity to describe here \emph{in extenso} what
quantum graphs are and why they are important; if such a need
nevertheless arises we can refer to papers from the dawns of the
history \cite{RS53}, from the times of new beginning in the
eighties \cite{GP88, ES89}, to more recent work containing a rich
bibliography \cite{KS99, Ku04}, and last not least, to the other
contributions making this volume.

As in the most of the mentioned work, the object of our interest
here are Schr\"odinger operators on metric graphs; we neglect
external fields and consider a free spinless particle on the
graph, with the Hamiltonian which acts as $H\psi_j= -\psi''_j$,
where $\psi_j$ denotes the wave function at the $j$th edge. It is
known for longtime \cite{ES89} that in order to make $H$
self-adjoint, a vertex joining $n$ graph edges may be
characterized by boundary conditions involving $n^2$ real
parameters; they have the form of a linear relation between
$\Psi(0)$, the column vector of the boundary values at the vertex
(identified conventionally with the origin of the coordinates),
and $\Psi'(0)$, the vector of the derivatives, taken all in the
outgoing direction.

A general and elegant form of these boundary conditions was found
in \cite{KS99}: any vertex in which $n$ edges meet can be
described by a pair of $n\times n$ matrices $A,B$ such that
$\mathrm{rank\,}(A,B) =n$ and the product $AB^*$ is self-adjoint.
The self-adjointness is guaranteed if the corresponding boundary
values satisfy the condition
 % ------------- %
 \begin{equation} \label{KS bc}
 A\Psi(0)+B\Psi'(0)=0\,.
 \end{equation}
 % ------------- %
Moreover, soon after several authors \cite{FT00, Ha00, KS00}
pointed out that the matrix pair in (\ref{KS bc}) can be made
unique by choosing
 % ------------- %
 \begin{equation} \label{AB through U}
 A=U-I\,,\quad B=i(U+I)
 \end{equation}
 % ------------- %
with a unitary $U$; a nontrivial coupling between the edges
corresponds naturally to the situation when the matrix $U$ is
non-diagonal. A simple proof of this fact for $n=2$ was given in
\cite{FT00} and extended to any $n$ in \cite{CE04}.

While the conditions (\ref{KS bc}) ensure self-adjointness of
quantum graph Hamiltonians, or in physical terms conservation of
probability current in the vertex, they say nothing about a
\emph{physical meaning of such a vertex coupling}. A natural way
to address the last question is to investigate approximations of a
quantum graph by more realistic systems with no free parameters.
An example is a quantum particle living in a configuration space
in the form of a thin tube-like domain; one can consider a family
of such domains shrinking to the given graph. A solution to this
problem at the level of eigenvalue convergence was found
\cite{KZ01, RS01, Sa01, EP05} in the situation that the tube-like
domain supports Laplacian with Neumann boundary conditions (or
similar operators), and an extension to the resolvent convergence
has been announced \cite{Po05}. These results, however, gave a
partial answer to the problem stated above because the limit leads
to the \emph{free boundary conditions},
 % ------------- %
 \begin{equation} \label{free}
 \psi_j(0)=\psi_k(0)\,,\; j,k=1,\dots,n\,, \quad \sum_{j=1}^n
 \psi'_j(0) = 0\,,
 \end{equation}
 % ------------- %
only. It is hoped that other approximating families, say, using
Dirichlet Laplacians, geometrically induced and/or external
potentials, could yield different vertex couplings, but this
problem is difficult and no such results are known at present.

A less ambitious program aims at approximating vertex couplings by
means of Schr\"odinger operators on the graph itself, using
suitable families of scaled potential, regular or singular. This
is relatively easy as long as we attempt to approximate couplings
with wavefunctions continuous at the vertex, i.e. the
one-parameter family of the so-called $\delta$ \emph{couplings}
\cite{Ex96a} described by the conditions
 % ------------- %
 \begin{equation} \label{delta}
 \psi_j(0)=\psi_k(0)=:\psi(0)\,,\; j,k=1,\dots,n\,, \quad \sum_{j=1}^n
 \psi'_j(0) = \alpha \psi(0)
 \end{equation}
 % ------------- %
with $\alpha\in\mathbb{R}$, which is obtained from (\ref{KS bc})
and (\ref{AB through U}) by choosing $U= {2\over n+i\alpha} J-I$,
where $J$ and $I$ are the $n\times n$ matrix whose all entries are
equal to one and the unit matrix, respectively. The procedure is
analogous to the approximation of $\delta$ interaction of the line
\cite{AGHH}: one starts with the conditions (\ref{free}) and adds
at each edge a naturally scaled potential, $V_{j,\epsilon}(x) =
\frac 1\epsilon V_j\left( \frac x\epsilon \right)$ for some
$V_j\in L^1$. A norm-resolvent limit then yields the $\delta$
coupling with $\alpha:= \sum_j \int V_j(x)\, \mathrm{d}x\:$
\cite{Ex96b}.

The situation is more complicated if the wavefunctions are
discontinuous at the vertex. The simplest example of such a
situation is the so-called $\delta'_s$ \emph{coupling},
 % ------------- %
 \begin{equation} \label{delta'_s}
 \psi'_j(0)=\psi'_k(0)=:\psi'(0)\,,\; j,k=1,\dots,n\,, \quad \sum_{j=1}^n
 \psi_j(0) = \beta \psi'(0)
 \end{equation}
 % ------------- %
with $\beta\in\mathbb{R}$. An inspiration can be found in the way
in which Cheon and Shigehara \cite{CS98a, CS98b} approximated
formally the $\delta'$ interaction on the line using a nonlinearly
scaled family of $\delta$ interactions -- their argument was later
shown to yield a norm-resolvent convergence and to lead to
approximations in terms of regular potentials \cite{AN00, ENZ01}.
It was shown in \cite{CE04}\footnote{A small correction is due
with respect to this paper: in the statement and proof of
Theorem~1 $\beta/n$ should be replaced everywhere by $\beta$.}
that the CS-type method can be used to approximate the $\delta'_s$
coupling for any $n$, the approximating operator domains having
functions continuous at the vertex. The aim of this paper is to
show that \emph{this result can be extended to couplings} with
discontinuous wavefunctions \emph{which are invariant with respect
to permutations of the graphs edges}: an approximation using a
$\delta$ coupling at the vertex and an $n$-tuple of $\delta$
interactions at the edges approaching the vertex will be derived
for all such couplings\footnote{We will be concerned with
nontrivial couplings only corresponding to non-diagonal matrices
$U$; in the opposite case the task is reduced to a much simpler
halfline problem -- cf.~\cite{FCT02}.}. We will see that in the
generic case the idea of \cite{CS98a} has a direct, albeit rather
tedious extension to the graph case, while for two one-parameter
subfamilies the choice of coupling parameters requires a
modification. Extensions to more general boundary conditions
inspired by \cite{SMC99} and approximations by regular potentials
are left to a subsequent publication.

\section{Permutation symmetric vertex couplings}

As in the previous work cited above we consider a \emph{star
graph} $\Gamma$ consisting of $n$ halflines meeting at a single
vertex. The corresponding Hilbert space is thus $\mathcal{H}=
\bigoplus_{j=1}^n L^2(\mathbb{R}_+)$. A general Hamiltonian
describing a free particle living on the graph is a self-adjoint
extension of the operator $H_0$ acting as $H_0\psi_j= -\psi''_j$
on functions $\Psi=\{\psi_j\}\in \bigoplus_{j=1}^n
W^{2,2}(\mathbb{R}_+)$ satisfying the conditions $\psi_j(0)=
\psi'_j(0)=0\:$; each such extension is specified by a boundary
condition (\ref{KS bc}) at the vertex.

Since the action of these operators at each component of the
wavefunction is the same, symmetry properties of the extensions
are given by those of the boundary conditions. We will be
interested in the permutation-invariant extensions, first
introduced in \cite{ES89}, which form a two-parameter family.
 % ------------- %
 \begin{proposition} \label{perm_inv}
 The boundary conditions (\ref{KS bc}) are permutation invariant
 if and only if the matrix $U$ in (\ref{AB through U}) equals
 % ------------- %
 \begin{equation}\label{tvar U}
 U=aI+bJ
 \end{equation}
 % ------------- %
 with complex coefficients $a,b$ satisfying the relations
 % ------------- %
 \begin{equation}\label{podm a,b}
 |a|=1 \quad \textrm{and} \quad |a+nb|=1.
 \end{equation}
 % ------------- %
 \end{proposition}
 % ------------- %
 \begin{proof}
The condition (\ref{KS bc}) is permutation invariant \emph{iff} it
is satisfied at the same time by the vectors $P\Psi(0)$ and
$P\Psi'(0)$ for any $P\in\mathcal{S}_n$. Multiplying it by
$P^{-1}$ from the left we get
 % ------------- %
 $$
 (P^{-1}UP-I)\Psi(0)+i(P^{-1}UP+I)\Psi'(0)=0\,,
 $$
 % ------------- %
the matrix $P^{-1}UP$ being obviously unitary. In view of the
uniqueness of the parametrization (\ref{AB through U}) the
property is equivalent to $P^{-1}UP=U$. Next we notice that a
simultaneous permutation of the rows and columns leaves the
diagonal elements on the diagonal, and the off-diagonal ones off
the diagonal; since $P^{-1}UP=U$ has to be satisfied for any
$P\in\mathcal{S}_n$ it follows that $U=aI+bJ$ for some $a,b\in
\mathbb{C}$. The conditions restricting the values of $a,b$ follow
from the unitarity of $U$,
 % ------------- %
 $$
 \left(UU^*\right)_{ij} =
 |a|^2\delta_{ij}+2\Re(a\bar{b})+n|b|^2 = \delta_{ij}\,,
 $$
 % ------------- %
which yields the relations
 % ------------- %
 $$
 |a|^2+2\Re(a\bar{b})+n|b|^2=1\,, \quad 2\Re(a\bar{b})+n|b|^2=0\,,
 $$
 % ------------- %
for $i=j$ and $i\ne j$, respectively. Substituting from the second
to the first one we get $|a|^2=1$. Finally, using $|a+nb|^2=|a|^2
+2n\Re(a\bar{b}) +n^2|b|^2$ we see that the left-hand side of the
second relation is a multiple of $|a+nb|^2-|a|^2$.
 \end{proof}

For definiteness we will denote in the following the self-adjoint
extension corresponding to fixed $a,b$ as $H^{a,b}$. Notice that
the boundary conditions described by Proposition~\ref{perm_inv}
can be also written more explicitly as the following system,
 % ------------- %
 $$
 (a-1)\psi_j(0)+b\sum^{n}_{k=1}\psi_k(0)+i(a+1)\psi_j'(0)+i
 b\sum^{n}_{k=1}\psi_k'(0)=0\,,\quad j=1,\dots,n\,,
 $$
 % ------------- %
which shows, in particular, that $(a-1)\psi_j(0)
+i(a+1)\psi_j'(0)$ is independent of $j$. To get a useful
equivalent formulation we subtract the $k$th one of these
condition from the $j$th one obtaining
 % ------------- %
 \begin{equation}\label{odectenim}
 (a-1)\left(\psi_j(0)-\psi_k(0)\right)+i(a+1)\left(\psi_j'(0)
 -\psi_k'(0)\right)=0\,, \quad j,k=1,\dots,n\,,
 \end{equation}
 % ------------- %
while summing all of them gives
 % ------------- %
 \begin{equation}\label{sectenim}
 (a-1+nb)\sum^{n}_{k=1}\psi_k(0)
 +i(a+1+nb)\sum^{n}_{k=1}\psi_k'(0)=0\,.
 \end{equation}
 % ------------- %
\begin{examples} \label{coupl ex}
We have already mentioned that $a=-1$ and $b=\frac{2}{n+i\alpha}$
describes the $\delta$ coupling (\ref{delta}), similarly $a=1$ and
$b=\frac{2}{i\beta-n}$ corresponds to the $\delta'_s$ coupling
(\ref{delta'_s}). Another example is the $\delta'$ coupling
\cite{Ex96a},
 % ------------- %
 \begin{equation} \label{delta'}
 \sum_{j=1}^n \psi'_j(0)=0\,,\quad
 \psi_j(0)-\psi_k(0) =
 {\beta\over n}
 (\psi'_j(0)-\psi'_k(0))\,,\; j,k=1,\dots,n\,,
 \end{equation}
 % ------------- %
referring to $a=\frac{i\beta+n}{i\beta-n}$ and
$b=\frac{2}{n-i\beta}$, and its dual counterpart $\delta_p$ with
the roles of functions and derivatives interchanged for which
$a=\frac{n-i\alpha}{n+i\alpha}$ and $b=-\frac{2}{n+i\alpha}$.

\end{examples}

\section{Approximation: a heuristic argument}

Let us describe the family we will employ to approximate
permutation-symmet\-ric Hamiltonians $H^{a,b}$. Let us recall that
we will consider all operators of this class with the exception of
those with a $\delta$ coupling, i.e. with the wavefunctions
continuous at the vertex, because for the latter we have the
natural approximation described in the introduction. We denote by
$H_{u,v}(d)$ the operator which is obtained from $H_{u,0}:=
H^{-1,2/(n+iu)}$ by adding a $\delta$ interaction of strength $v$
to each edge at the distance $d$ from the vertex; it is the same
scheme which was used in the particular case of $\delta'_s$
treated in \cite{CE04}. The aim of the present section is to
derive formally how the values of the parameters $u,v$ as
functions of $d$ should be chosen.

At the vertex the boundary condition defining $H_{u,v}(d)$ are of
the form (\ref{delta}) with $\alpha$ replaced by $u$, while the
added $\delta$ interactions are characterized by
 % ------------- %
 \begin{equation}\label{II.serie}
 \psi_j(d+)=\psi_j(d-)=:\psi_j(d)\,, \;\;
 \psi_j'(d+)-\psi_j'(d-)=v\psi_j(d)\,,\quad j=1,\dots,n\,.
 \end{equation}
 % ------------- %
To find relations between the boundary values, we employ Taylor
expansion
 % ------------- %
 \begin{equation}\label{III.serie}
 \psi_j(d)=\psi_j(0)+d\psi_j'(0)+\mathcal{O}(d^2)\,, \;\;
 \psi_j'(d-)=\psi_j'(0+)+\mathcal{O}(d)\,, \quad j=1,\dots,n\,;
 \end{equation}
 % ------------- %
we want to choose $u,v$ to get the relations \eqref{odectenim} and
\eqref{sectenim} in the limit $d\to 0+$. The first one of the
relations (\ref{III.serie}) together with the continuity at the
vertex imply
 % ------------- %
 \begin{equation}\label{prvniuprava}
 \psi_j(d)-\psi_k(d)=d\left(\psi_j'(0)
 -\psi_k'(0)\right)+\mathcal{O}(d^2)\,.
 \end{equation}
 % ------------- %
Furthermore, the second one of the relations (\ref{III.serie}) in
combination with (\ref{II.serie}) tell us that the difference
$\psi_j'(0+)-\psi_k'(0+)$ is equal to
 % ------------- %
 \begin{equation}\label{druhauprava}
 \psi_j'(d-)-\psi_k'(d-)+\mathcal{O}(d)
 =\psi_j'(d+)-\psi_k'(d+)-v(\psi_j(d)
 -\psi_k(d))+\mathcal{O}(d)\nonumber
 \end{equation}
 % ------------- %
giving thus $d\left(\psi_j'(d+)-\psi_k'(d+)) -v(\psi_j(d)
-\psi_k(d))\right) +\mathcal{O}(d^2)$ as the value of the
left-hand side in \eqref{prvniuprava}, which can be rewritten as
 % ------------- %
 \begin{equation}\label{ctvrtauprava}
(1+dv)\left(\psi_j(d)-\psi_k(d)\right)-d\left(\psi_j'(d+)-\psi_k'(d+)\right)
=\mathcal{O}(d^2)\,.\nonumber
 \end{equation}
 % ------------- %
This should give $(a-1)(\psi_j(0)-\psi_k(0)) +i(a+1)(\psi_j'(0+)
-\psi_k'(0+))=0$ in the limit $d\to 0+$. As we have mentioned
above, the case of a $\delta$ interaction in which we have $a=-1$
is excluded, hence we are allowed to require $\frac{1+dv}{d}
=\frac{a-1}{i(a+1)}$. This in turn yields the following relation
for the parameter $v$,
 % ------------- %
\begin{equation}\label{v}
v=-\frac{1}{d}-i\frac{a-1}{a+1}\,;
\end{equation}
 % ------------- %
notice that it is real-valued in view of the condition $|a|=1$,
because
 % ------------- %
 $$
 i\frac{a-1}{a+1}=
 i\frac{|a|^2+2i\Im a-1}{|a+1|^2}
 =-2\frac{\Im a}{|a+1|^2}\in\mathbb{R}\,.
 $$
 % ------------- %
It remains to find $u$. We employ again the first of the relations
\eqref{III.serie} together with both the vertex conditions
(\ref{delta}) for $\alpha=u$ rewriting in this way
$\sum^{n}_{j=1}\psi_j(d)$ as
 % ------------- %
 \begin{equation}\label{prvniupravab}
 n\psi(0) +d\sum^{n}_{j=1}\psi_j'(0+)+\mathcal{O}(d^2)
 =\frac{n}{u}\sum^{n}_{j=1}\psi_j'(0+)
 +d\sum^{n}_{j=1}\psi_j'(0+)+\mathcal{O}(d^2) \nonumber
 \end{equation}
 % ------------- %
As before we use \eqref{II.serie} and \eqref{III.serie} to
eliminate $\psi_j'(0+)$,
 % ------------- %
 \begin{equation}\label{druhaupravab}
 \sum^{n}_{j=1}\psi_j'(0+)=\sum^{n}_{j=1}\psi_j'(d-)+\mathcal{O}(d)
 =\sum^{n}_{j=1}\psi_j'(d+)-v\sum^{n}_{j=1}\psi_j(d)+\mathcal{O}(d)\nonumber
 \end{equation}
 % ------------- %
Substituting into the expression for $\sum^{n}_{j=1}\psi_j(d)$ we
get after a simple manipulation
 % ------------- %
 \begin{equation}\label{ctvrtaupravab}
 \left(1+v\left(\frac{n}{u}+d\right)\right)\sum^{n}_{j=1}\psi_j(d)
 =\left(\frac{n}{u}+d\right)\left(\sum^{n}_{j=1}\psi_j'(d+)
 +\mathcal{O}(d)\right)+\mathcal{O}(d^2)\,.\nonumber
 \end{equation}
 % ------------- %
using the value of $v$ given by \eqref{v} we find that the
quantity
 % ------------- %
 \begin{equation}\label{sestaupravab}
 \left(\left(\frac{1}{d}+i\frac{a-1}{a+1}\right)\frac{n}{u}
 +i\frac{a-1}{a+1}d\right)\sum^{n}_{j=1}\psi_j(d)
 +\left(\frac{n}{u}+d\right)\left(\sum^{n}_{j=1}\psi_j'(d+)
 +\mathcal{O}(d)\right)
 \end{equation}
 % ------------- %
behaves as $\mathcal{O}(d^2)$ in the limit $d\to 0+$. We will look
for $u$ having a stronger singularity than $v$ assuming
$\frac{1}{u}=\mathcal{O}(d^2)$; then the last claim simplifies as
follows,
 % ------------- %
 \begin{equation}\label{sedmaupravab}
 \left(\frac{1}{d^2}\frac{n}{u}
 +i\frac{a-1}{a+1}\right)\sum^{n}_{j=1}\psi_j(d)
 +\sum^{n}_{j=1}\psi_j'(d+)=\mathcal{O}(d)\,.\nonumber
 \end{equation}
 % ------------- %
This is required to give the condition (\ref{sectenim}) in the
limit $d\to 0+$ which happens if
 % ------------- %
 $$
 \frac{1}{d^2}\frac{n}{u}+i\frac{a-1}{a+1}
 =\frac{a-1+nb}{i(a+1+nb)}\,,
 $$
 % ------------- %
provided the two denominators containing the coupling parameters
do not vanish. The first one is zero for the $\delta$ coupling
which we have excluded from the outset, the second one vanishes
\emph{iff} $a,b$ correspond to the $\delta_p$ coupling described
in Examples~\ref{coupl ex}. It is also clear that in view of the
conditions $|a|=1$ a $|a+nb|=1$ the fractions $\frac{a-1+nb}
{a+1+nb}$ and $\frac{a-1}{a+1}$ are purely imaginary. This
motivates us to choose
 % ------------- %
 \begin{equation}\label{u}
 u=i\frac{n}{d^2} \left(\frac{a-1+nb}{a+1+nb}
 +\frac{a-1}{a+1}\right)^{-1}
 \end{equation}
 % ------------- %
assuming that the expression in the parentheses is nonzero which
is true as long as
 % ------------- %
 \begin{equation}\label{one more}
  a(a+nb)\ne 1\,.
 \end{equation}
 % ------------- %
The parameter $u$ defined by (\ref{u}) is, of course, real and
$u=\mathcal{O}(d^{-2})$ as $d\to 0+$; this concludes our search
for the approximating operator family in the generic case.

It remains to carry on the heuristic argument for the two excluded
one-parame\-ter subfamilies, the $\delta_p$ coupling and the one
violating the condition (\ref{one more}). We will show that the
coupling of the $\delta$ interactions at the graph arms can be
preserved, it is only necessary to change the function $u$
describing the vertex. Let us first suppose that the latter has a
stronger singularity at $d=0$, for instance,
 % ------------- %
 \begin{equation} \label{u1}
 u=\frac{\zeta}{d^3}
 \end{equation}
 % ------------- %
for a fixed nonzero $\zeta\in\mathbb{R}$ (in fact, one can replace
$d^3$ by $d^\nu$ for any $\nu>2$). Substituting this into
(\ref{sestaupravab}) we get a condition which in the limit $d\to
0+$ yields
 % ------------- %
 $$
 i\frac{a-1}{a+1}\sum^{n}_{j=1}\psi_j(0)+\sum^{n}_{j=1}\psi_j'(0)=0\,.
 $$
 % ------------- %
The left-hand side makes sense since $a\ne -1$ and it is easy to
check that if (\ref{one more}) is not valid, i.e. $a+nb=a^{-1}$,
the last relation is equivalent to (\ref{sectenim}). On the other
hand, to deal with the $\delta_p$ coupling we take $u$ with a pole
singularity,
 % ------------- %
 \begin{equation} \label{u2}
 u=-\frac{n}{d}\,.
 \end{equation}
 % ------------- %
The second term in (\ref{sestaupravab}) then vanishes and we find
that $\sum^{n}_{j=1}\psi_j(0)=\mathcal{O}(d^2)$ which gives in the
limit $d\to 0+$ the condition (\ref{sectenim}) for the particular
case of $\delta_p$.

\section{The main result}

Now we are ready to formulate and prove our main result.
 % ------------- %
\begin{theorem}
Given complex numbers $a\ne -1$ and $b\ne 0$ satisfying the
conditions (\ref{podm a,b}), define $u=u(d)$ and $v=v(d)$ for
$d>0$ as in the previous section, i.e. by the relations \eqref{u},
\eqref{u1}, \eqref{u2}, and \eqref{v}, respectively; then the
operators $H_{u,v}(d)$ converge to $H^{a,b}$ in the norm resolvent
topology as $d\to 0+$.
\end{theorem}
 % ------------- %
\begin{proof}
To begin with we observe that the permutation symmetry of the
boundary conditions \eqref{odectenim} and \eqref{sectenim} allows
us to simplify the task by reducing it to independent halfline
problems. To this aim let us find the spectrum of the matrix $U$.
Since the latter equals $U=aI+bJ$ it is sufficient to look at the
matrix $J$ which has rank one, and thus zero is its eigenvalue of
multiplicity $n-1$ corresponding to vectors with vanishing
component sum; the remaining simple eigenvalue is $n$. The
corresponding eigenvalues of $U$ are $a$ and $a+nb$, respectively,
with the same multiplicities. Consider first the generic case
where these eigenvalues are not inverse to each other by (\ref{one
more}) and none of them equals to $-1$.

The symmetry allows us to decompose the operator in question,
$H^{a,b}$ on the Hilbert space $\mathcal{H} =\bigoplus_{j=1}^{n}
L^2(\mathbb{R}^+) =L^2(\mathbb{R}^+)\otimes \mathbb{C}^n$, into
orthogonal sum of two components. The first one denoted as
$H^{(1)a,b}$ acts at the ``scalar'' subspace isomorphic to
$L^2(\mathbb{R}^+)\otimes\mathbb{C}$ consisting of functions
$\Psi\in\mathcal{H}$ which are symmetric with respect to
permutations, $\psi_j(x) =\psi_k(x)$ for all $j,k=1,\dots,n$; it
is characterized by the boundary conditions
$(a+nb-1)\Psi(0)+i(a+nb+1)\Psi'(0)=0$. The other one for which we
use the symbol $H^{(n-1)a,b}$ acts on the orthogonal complement
which is isomorphic to $L^2(\mathbb{R}^+)\otimes
\mathbb{C}^{n-1}$ consisting of $\Psi\in\mathcal{H}$ with
vanishing component sum. The action of $H^{(n-1)a,b}$ on all
linear combinations $\sum^{n}_{j=1}c_j\psi_j(x)$ is identical and
the boundary conditions are $(a-1)\Psi(0)+i(a+1)\Psi'(0)=0$.

In the same way one can decompose the approximating operators
$H_{u,v}(d)$. The part $H^{(1)}_{u,v}(d)$ acts on the ``scalar''
subspace of functions invariant with respect to permutations, the
boundary conditions being $\Psi'(0)= \frac un \Psi(0)$. The
remaining component acts on ``(n-1)-dimensional vector functions''
being isomorphic to $n-1$ copies of the ``scalar'' problem with
Dirichlet boundary conditions.

We will use the fact that the resolvents of all the involved
operators can be constructed explicitly using a standard ODE
result in combination with Krein's formula. Let us consider first
the part independent of the coupling at the vertex. For a fixed
$k$ from the upper complex halfplane the Green function of the
Laplacian on the halfline with Dirichlet condition at the origin
is
 % ------------- %
 $$
 \mathcal{G}_k(x,y)=\frac{1}{k}\sin(kx_<)\mathbb{e}^{i kx_>}
 =\frac{1}{\kappa}\sinh(\kappa x_<)\mathbb{e}^{-\kappa x_>}\,,
 $$
 % ------------- %
where we denote conventionally $x_<=\min\{x,y\}$,
$x_>=\max\{x,y\}$, and $\kappa=-ik$. The $\delta$ interaction at
the point $x=d$ represents a rank-one perturbation of the above
free resolvent, and corresponding Green's function is found easily
with the help of (\ref{II.serie}) as in \cite[Sec.~I.3]{AGHH} or
\cite{CE04} to be equal to
 % ------------- %
 \begin{equation}\label{aproximujici (n-1)-dim.}
 \mathcal{G}_k^v(x,y)=\mathcal{G}_k(x,y)
 +\frac{\mathcal{G}_k(x,d)\mathcal{G}_k(d,y)}
 {-v^{-1}-\mathcal{G}_k(d,d)}\,.
 \end{equation}
 % ------------- %
Next we have to find the Green function of the approximated
operator. Following \cite[Sec.~8.4]{We} we need a solution of the
equation $-\psi''=k^2\psi$ satisfying the condition
$(a-1)\psi(0)+i(a+1)\psi'(0)=0$ and its Wronskian with $\phi(x)=
\mathrm{e}^{-\kappa x}$; this yields
 % ------------- %
 \begin{equation}\label{aproximovana (n-1)-dim.}
 \mathcal{G}_{i\kappa}^{a}(x,y)
 =\frac{\left(i(a-1)\sinh\kappa x_<
 +\kappa(a+1)\cosh\kappa x_<\right)\mathrm{e}^{-\kappa x_>}}
 {\kappa(i(a-1)+\kappa(a+1))}\,.
 \end{equation}
 % ------------- %
Now we are going to show that $\mathcal{G}_{i\kappa}^{v}$
converges to $\mathcal{G}_{i\kappa}^{a}$ pointwise as $d\to 0+$.
First we suppose that both arguments are not smaller than $d$;
without loss of generality we may put $d\le x\leq y$ rewriting the
difference $\mathcal{G}_{i\kappa}^{v}(x,y)
-\mathcal{G}_{i\kappa}^{a}(x,y)$ as
 % ------------- %
 $$
 \frac{\mathrm{e}^{-\kappa y}}{\kappa}\left(
 \sinh\kappa x+ \frac{\mathrm{e}^{-\kappa x}\sinh^2\kappa d}
 {\frac{\kappa d(a+1)}{a+1+i d(a-1)}- \mathrm{e}^{-\kappa d}
 \sinh\kappa d}
 - \frac{i(a\!-\!1)\sinh\kappa x+\kappa(a\!+\!1)\cosh\kappa x}
 {i(a\!-\!1)+\kappa(a\!+\!1)}\right)
 $$
 % ------------- %
The sum of the first and the third term in the bracket equals
$\frac{-\kappa(a+1)\mathrm{e}^{-\kappa x}}{i(a-1)+\kappa(a+1)}$
being independent of $d$. In the second term we use the expansion
$\sinh(x)=x+\mathcal{O}(x^3)$ and obtain after simple
manipulations
 % ------------- %
 $$
 \frac{\mathrm{e}^{-\kappa x}\sinh^2\kappa d}
 {\frac{\kappa d(a+1)}{a+1+i d(a-1)}- \mathrm{e}^{-\kappa d}
 \sinh\kappa d}
 = \mathrm{e}^{-\kappa x}\left(\frac{\kappa(a+1)}
 {i(a-1)+\kappa(a+1)}+\mathcal{O}(d^2)\right)\,,
 $$
 % ------------- %
hence the terms non-vanishing in the limit cancel and there is a
$K>0$ such that
 % ------------- %
 $$
 \left|\mathcal{G}_{i\kappa}^{v}(x,y)
 -\mathcal{G}_{i\kappa}^{a}(x,y)\right|
 <K\,\mathrm{e}^{-\kappa x}
 \mathrm{e}^{-\kappa y}\:d^2
 $$
 % ------------- %
holds for all $d<x<y$, and by the same argument also for $d<y<x$.

Next we suppose that $x\leq d\leq y$ when the Green's function
difference is
 % ------------- %
 $$
 \frac{\mathrm{e}^{-\kappa y}}{\kappa}\left(
 \sinh\kappa x+ \frac{\sinh\kappa x\,
 \mathrm{e}^{-\kappa d}\sinh^2\kappa d}
 {\frac{\kappa d(a+1)}{a+1+i d(a-1)}- \mathrm{e}^{-\kappa d}
 \sinh\kappa d}
 - \frac{i(a\!-\!1)\sinh\kappa x+\kappa(a\!+\!1)\cosh\kappa x}
 {i(a\!-\!1)+\kappa(a\!+\!1)}\right)
 $$
 % ------------- %
We want to show that the expression in the bracket is uniformly
bounded in $x,d$ provided $d$ is small enough; we may suppose that
$d<1$. The first and the third term are bounded in view of the
continuity at $x=0$, the middle one is easily found to be
$\mathcal{O}(1)$ as $d\to 0+$. Hence there is an $L>0$ independent
of $x,y$ and $d$ such that $\left|\mathcal{G}_{i\kappa}^{v}(x,y)
-\mathcal{G}_{i\kappa}^{a}(x,y)\right|<L\,
\mathrm{e}^{-\Re(\kappa)y}$, and in a similar way one can estimate
the Green function difference for $y\le d<x$. It remains to deal
with the case when both argument are less than $d$, say $x<y<d$; we
may again suppose that $d<1$. The middle term in the above
expression is then replaced by
 % ------------- %
 $$
 \frac{\sinh\kappa x\,\sinh\kappa y\,
 \mathrm{e}^{-2\kappa d}}
 {\frac{\kappa d(a+1)}{a+1+i d(a-1)}- \mathrm{e}^{-\kappa d}
 \sinh\kappa d}
 $$
 % ------------- %
and one checks easily that there is an $M>0$ independent of $x,y$
and $d$ such that the pointwise estimate
$\left|\mathcal{G}_{i\kappa}^{v}(x,y) -\mathcal{G}_{i\kappa}^{a}
(x,y)\right|<M$ holds.

These bounds allow us to estimate Hilbert-Schmidt norm of the
difference,
 % ------------- %
 \begin{eqnarray}
 \left\|R_{H^{(n-1)}_{u,v}(d)}(k^2)-R_{H^{(n-1)a,b}}(k^2)\right\|_2^2=
 \int^{\infty}_{0}\int^{\infty}_{0}\left|
 \mathcal{G}_{i\kappa}^{v}(x,y)
 -\mathcal{G}_{i\kappa}^{a}(x,y)\right|^2\mathrm{d}x\mathrm{d}y \nonumber\\
 =\int^{d}_{0}\int^{d}_{0}\left|\mathcal{G}_{i\kappa}^{v}(x,y)
 -\mathcal{G}_{i\kappa}^{a}(x,y)\right|^2\mathrm{d}x\mathrm{d}y+
 \int^{d}_{0}\int^{\infty}_{d}\left|\mathcal{G}_{i\kappa}^{v}(x,y)
 -\mathcal{G}_{i\kappa}^{a}(x,y)\right|^2\mathrm{d}x\mathrm{d}y \nonumber\\
 +\int^{\infty}_{d}\int^{d}_{0}\left|\mathcal{G}_{i\kappa}^{v}(x,y)
 -\mathcal{G}_{i\kappa}^{a}(x,y)\right|^2\mathrm{d}x\mathrm{d}y+
 \int^{\infty}_{d}\int^{\infty}_{d}\left|\mathcal{G}_{i\kappa}^{v}(x,y)
 -\mathcal{G}_{i\kappa}^{a}(x,y)\right|^2\mathrm{d}x\mathrm{d}y\nonumber
\end{eqnarray}
 % ------------- %
It is straightforward to check that last integral does not exceed
the value
 % ------------- %
 $$
 \int^{\infty}_{d}\int^{\infty}_{d}
 \left(\mathrm{e}^{-\Re(\kappa)x}\mathrm{e}^{-\Re(\kappa)y}
 Kd^2\right)^2\mathrm{d}x\mathrm{d}y \leq
 \left(\frac{K}{2\Re(\kappa)}\right)^2d^4\,,
 $$
 % ------------- %
and similarly the first one and the middle two are estimated by
$4M^2d^2$ and $\frac{L^2}{2\Re(\kappa)}\,d$, respectively, which
means that
 % ------------- %
 $$
 \lim_{d\to 0+}\left\|R_{H^{(n-1)}_{u,v}(d)}(k^2)
 -R_{H^{(n-1)a,b}}(k^2)\right\|_2^2=0\,,
$$
 % ------------- %
and the same is \emph{a fortiori} true for the operator norm. This
concludes the argument for the first component of the operator.

The proof for the ``scalar'' component is similar, just a bit more
complicated, so we can skip some details. First we construct the
Green function for the $\delta$ coupling with the parameter
$u\in\mathbb{R}$ projected on the subspace of functions with
coinciding components; in a similar way as above we find that the
resolvent kernel equals
 % ------------- %
 $$
 \mathcal{G}_{i\kappa}^{u}(x,y)=\frac{\mathrm{e}^{-\kappa x_>}}
 {\kappa\left(\frac{u}{n}+\kappa\right)}
 \left(\frac{u}{n}\sinh\kappa x_<
 +\kappa\cosh\kappa x_<\right)\,.
 $$
 % ------------- %
An analogous construction for the (negative) Laplacian with the
boundary condition $(a+nb-1)\psi(0)+i(a+nb+1)\psi'(0)=0$ at the
origin gives the Green function
 % ------------- %
 $$
 \mathcal{G}_{i\kappa}^{a,b}(x,y)=\frac{\mathrm{e}^{-\kappa y}\:
 \left(i(a+nb-1)\sinh\kappa x
 +\kappa(a+nb+1)\cosh\kappa x\right)}
 {\kappa(i(a+nb-1)+\kappa(a+nb+1))}\,.
 $$
 % ------------- %
Finally the resolvent kernel of the approximating function, to be
compared with the last expression, is obtained again from
$\mathcal{G}_{i\kappa}^{u}(x,y)$ by means of Krein's formula
 % ------------- %
 $$
 \mathcal{G}_{i\kappa}^{u,v}(x,y)
 =\mathcal{G}_{i\kappa}^u(x,y)
 +\frac{\mathcal{G}_{i\kappa}^u(x,d)\mathcal{G}_{i\kappa}^u(d,y)}
 {-v^{-1}-\mathcal{G}_{i\kappa}^u(d,d)}\,.
 $$
 % ------------- %
To estimate the Green function difference we assume again first
that $d\leq x\leq y$, so
 % ------------- %
\begin{eqnarray*}
\lefteqn{\mathcal{G}_{i\kappa}^{u,v}(x,y)
-\mathcal{G}_{i\kappa}^{a,b}(x,y)}
\\ && = \frac{\mathrm{e}^{-\kappa y}}{\kappa}
\Bigg( \frac{\frac{u}{n}\sinh\kappa x +\kappa\cosh\kappa
x}{\frac{u}{n}+\kappa} + \frac{\frac{\mathrm{e}^{-\kappa
x}}{\left(\frac{u}{n}
+\kappa\right)^2}\left(\frac{u}{n}\sinh\kappa d+\kappa\cosh\kappa
d\right)^2} {\frac{\kappa d(a+1)}{a+1+i d(a-1)}-
\frac{\mathrm{e}^{-\kappa d}}{\frac{u}{n}+\kappa}
\left(\frac{u}{n}\sinh\kappa d+\kappa\cosh\kappa d\right)} \\
&& \phantom{iii} - \frac{i(a+nb-1)\sinh\kappa
x+\kappa(a+nb+1)\cosh\kappa x}{i(a+nb-1)+\kappa(a+nb+1)}\Bigg)
\end{eqnarray*}
 % ------------- %
where we have used (\ref{v}) and $u$ should be substituted from
(\ref{u}). Our aim is to find the behavior of this expression for
small $d$ using the expansion
 % ------------- %
 $$
 \cosh(x)=1+\mathcal{O}(x^2)\,,\quad
 \frac{1}{1+x}=1-x+\mathcal{O}(x^2) \quad
 \mathrm{as} \;\; x\to 0.
 $$
 % ------------- %
The first term gives
 % ------------- %
 $$
 \frac{\frac{u}{n}\sinh\kappa x +\kappa\cosh\kappa x}
 {\frac{u}{n}+\kappa} =\sinh\kappa x+\mathrm{e}^
 {-\kappa x}\: \mathcal{O}(d^2)\,,
 $$
 % ------------- %
for the second one we get after a straightforward but tedious
computation
 % ------------- %
 $$
 \mathrm{e}^{-\kappa x}
 \left(\frac{\kappa(a+1+nb)}{\kappa(a+1+nb)+i(a-1+nb)}
 +\mathcal{O}(d)\right)\,,
 $$
 % ------------- %
and the third is independent of $d$; putting everything together
we find
 % ------------- %
 $$
 \mathcal{G}_{i\kappa}^{u,v}(x,y)
 -\mathcal{G}_{i\kappa}^{a,b}(x,y)
 =\mathrm{e}^{-\kappa x}\mathrm{e}^{-\kappa y}\:\mathcal{O}(d)\,,
 $$
 % ------------- %
because the non-vanishing terms cancel again. In other words,
there is a $K'>0$ independent of $x,y$ and $d$ such that the
following inequality
 % ------------- %
 \begin{equation} \label{both larger}
 \left|\mathcal{G}_{i\kappa}^{u,v}(x,y)
 -\mathcal{G}_{i\kappa}^{a,b}(x,y)\right| <
 K'\mathrm{e}^{-\kappa x}\mathrm{e}^{-\kappa y}\:d
 \end{equation}
 % ------------- %
holds. In the ``mixed'' case, $x\le d\le y$, we have
 % ------------- %
 \begin{eqnarray*}
 \lefteqn{\mathcal{G}_{i\kappa}^{u,v}(x,y)
 -\mathcal{G}_{i\kappa}^{a,b}(x,y)=
 \frac{\mathrm{e}^{-\kappa y}}{\kappa}\left(
 \frac{\frac{u}{n}\sinh\kappa x+\kappa\cosh\kappa x}
 {\frac{u}{n}+\kappa}+\right.} \\ &&
 +\frac{\frac{\mathrm{e}^{-\kappa d}}
 {\left(\frac{u}{n}+\kappa\right)^2}
 \left(\frac{u}{n}\sinh\kappa x
 +\kappa\cosh\kappa x\right)\left(\frac{u}{n}\sinh\kappa d
 +\kappa\cosh\kappa d\right)} {\kappa\frac{d(a+1)}{a+1+i d(a-1)}-
 \frac{\mathrm{e}^{-\kappa d}}{\frac{u}{n}+\kappa}
 \left(\frac{u}{n}\sinh\kappa d+\kappa\cosh\kappa d\right)} \\
 && - \left.\frac{i(a+nb-1)\sinh\kappa x+\kappa(a+nb+1)\cosh\kappa x}
 {i(a+nb-1)+\kappa(a+nb+1)}\right)\,.
\end{eqnarray*}
 % ------------- %
The first and the third term at the right-hand side are obviously
bounded independently of $x,y$ and $d$, and in the same way as
above one can check that the second one is $\mathcal{O}(1)$ as
$d\to 0+$, hence there is an $L'$ independent of $x,y$ and $d<1$
such that
 % ------------- %
 $$
 \left|\mathcal{G}_{i\kappa}^{u,v}(x,y)
 -\mathcal{G}_{i\kappa}^{a,b}(x,y)\right| <
 \mathrm{e}^{-\Re(\kappa)y}\,L'\,.
 $$
 % ------------- %
The same is naturally true if the roles of $x$ and $y$ are
interchanged. It remains to analyze the situation when both $x,y$
do not exceed $d$, say $x\le y\le d$, when
 % ------------- %
 \begin{eqnarray*}
 \lefteqn{\mathcal{G}_{i\kappa}^{u,v}(x,y)
 -\mathcal{G}_{i\kappa}^{a,b}(x,y)=
 \frac{\mathrm{e}^{-\kappa y}}{\kappa}\,
 \frac{\frac{u}{n}\sinh\kappa x+\kappa\cosh\kappa x}
 {\frac{u}{n}+\kappa}} \\ &&
 +\frac{\frac{\mathrm{e}^{-2\kappa d}}
 {\kappa\left(\frac{u}{n}+\kappa\right)^2}
 \left(\frac{u}{n}\sinh\kappa x
 +\kappa\cosh\kappa x\right)\left(\frac{u}{n}\sinh\kappa y
 +\kappa\cosh\kappa y\right)} {\frac{\kappa d(a+1)}{a+1+i d(a-1)}-
 \frac{\mathrm{e}^{-\kappa d}}{\frac{u}{n}+\kappa}
 \left(\frac{u}{n}\sinh\kappa d+\kappa\cosh\kappa d\right)} \\
 && - \frac{\mathrm{e}^{-\kappa y}}{\kappa}\,
 \frac{i(a+nb-1)\sinh\kappa x+\kappa(a+nb+1)
 \cosh\kappa x}{i(a+nb-1)+\kappa(a+nb+1)}\,.
\end{eqnarray*}
 % ------------- %
In the same way as above one establishes existence of an $M'>0$
independent of $x,y$ and $d<1$ such that
 % ------------- %
 $$
 \left|\mathcal{G}_{i\kappa}^{u,v}(x,y)
 -\mathcal{G}_{i\kappa}^{a,b}(x,y)\right| < M'\,.
 $$
 % ------------- %
Using these bounds and repeating the above Hilbert-Schmidt
estimate we get
 % ------------- %
 $$
 \lim_{d\to 0+}\left\|R_{H^{(1)}_{u,v}(d)}(k^2)
 -R_{H^{(1)a,b}}(k^2)\right\|_2^2=0\,,
 $$
 % ------------- %
which implies the analogous limiting relation for the operator
norm of the resolvent difference which we set out to prove.

In the remaining two cases it is sufficient to consider the
``scalar'' component because the orthogonal complement does not
contain the parameter $u$. Take first the case when the condition
(\ref{one more}) is violated. If the variables satisfy $d\leq
x\leq y$ we can rewrite the Green function difference using
(\ref{u1}) as
 % ------------- %
\begin{eqnarray*}
\lefteqn{\mathcal{G}_{i\kappa}^{u,v}(x,y)
-\mathcal{G}_{i\kappa}^{a,b}(x,y) = \frac{\mathrm{e}^{-\kappa
y}}{\kappa} \Bigg( \frac{\frac{\zeta}{n}\sinh\kappa x +\kappa
d^3\cosh\kappa x}{\frac{\zeta}{n}+\kappa d^3}} \\ &&
\phantom{AAAAA} +\frac{\frac{\mathrm{e}^{-\kappa
x}}{\left(\frac{\zeta}{n} +\kappa
d^3\right)^2}\left(\frac{\zeta}{n}\sinh\kappa d+\kappa
d^3\cosh\kappa d\right)^2} {\frac{\kappa d(a+1)}{a+1+i d(a-1)}-
\frac{\mathrm{e}^{-\kappa d}}{\frac{\zeta}{n}+\kappa d^3}
\left(\frac{\zeta}{n}\sinh\kappa d+\kappa d^3\cosh\kappa d\right)}  \\
&& \phantom{AAAAA} - \frac{i(a+nb-1)\sinh\kappa
x+\kappa(a+nb+1)\cosh\kappa x}{i(a+nb-1)+\kappa(a+nb+1)}\Bigg)\,;
\end{eqnarray*}
 % ------------- %
expanding the first two terms at the right-hand side we establish
existence of a $K'>0$ independent of $x,y$ and $d$ such that the
inequality (\ref{both larger}) holds. In a similar way one
proceeds when one or both arguments are smaller than $d$. The same
can be done in the $\delta_p$ case where the resolvent difference
for $d\leq x\leq y$ is
 % ------------- %
\begin{eqnarray*}
\lefteqn{\mathcal{G}_{i\kappa}^{u,v}(x,y)
-\mathcal{G}_{i\kappa}^{a,b}(x,y) = \frac{\mathrm{e}^{-\kappa y}}
{\kappa} \Bigg( \frac{-\sinh\kappa x+ \kappa d\cosh\kappa x}
{-1+\kappa d}} \\ && + \frac{\frac{\mathrm{e}^{-\kappa x}}
{(-1+\kappa d)^2} \left(-\sinh\kappa d+\kappa d\cosh\kappa
d\right)^2} {\kappa\frac{d(a+1)}{a+1+i d(a-1)}-
\frac{\mathrm{e}^{-\kappa d}} {-1+\kappa d} (-\sinh\kappa d
+\kappa d\cosh\kappa d)} -\sinh\kappa x\Bigg)
\end{eqnarray*}
 % ------------- %
and the other variable combinations are dealt with analogously.
The Hilbert-Schmidt estimate is the same as in the generic case;
this concludes the proof. \end{proof}

\section{Concluding remarks} \label{concl}

We have mentioned in the introduction that approximation including
singular couplings can be used an intermediate step in a search
for approximations based on regular potentials. In this sense
$\delta$ coupling and $\delta$ interactions are preferable because
in this case we already know how to make the second step; hence
our result paves way to a complete potential approximation of
permutation symmetric couplings.

In particular, comparing with \cite{CE04} we do not need
$\delta_p$ coupling to approximate $\delta'$, and the $\delta_p$
itself can be approximated by $\delta$ interactions.  We have
seen, however, that in this case the central singularity is of a
pole type with respect to $d$ similarly as the couplings of the
$\delta$'s at graph edges. This illustrates the exceptional
character of $\delta_p$ which is in a sense akin to $\delta$, with
the roles of the ``scalar'' and $(n-1)$-components interchanged.
The remaining one-parameter family of couplings violating the
condition (\ref{one more}) needs, on the contrary, a stronger
singularity with respect to $d$ at the vertex. The reason of this
behavior is not clear; this underlines one more time the fact that
our present understanding to the zoology of vertex couplings in
quantum graphs is still far from satisfactory.

\bibliographystyle{amsalpha}

\end{document}